\documentclass[10pt,twocolumn,showpacs,prl]{revtex4}

\usepackage{graphicx}
\usepackage{mathptmx}

\newcommand{\J}[2]{J_{#1, #2}}

\renewcommand{\ss}{\mathbf{\sigma}}
\newcommand{\s}{\sigma}

\newcommand{\bfigref}[1]{Figure~\ref{#1}}

\newcommand{\figref}[1]{Fig.~\ref{#1}}

\newcommand{\defeq}{:=}

\newcommand{\Prob}[1]{\mathrm{Prob}\{#1\}}
\newcommand{\eqref}[1]{(\ref{#1})}

\begin{document}

\title{Competitive nucleation and the Ostwald rule\\ in a generalized Potts
model
with multiple metastable phases}

\author{David P. Sanders}
\email{dsanders@fis.unam.mx}
\author{Hern\'an Larralde}
\author{Fran\c{c}ois Leyvraz}
\affiliation{\mbox{Instituto de Ciencias F\'{\i}sicas, Universidad Nacional
Aut\'onoma
de M\'exico, Apartado postal 48-3, 
62551 Cuernavaca, Morelos, Mexico}}

\begin{abstract}
We introduce a simple nearest-neighbor spin model with multiple
metastable phases, the number and decay pathways of which are
explicitly controlled by the parameters of the system. With this model
we can construct, for example, a system which evolves through an
arbitrarily long succession of metastable phases. We also
construct systems in which different phases may nucleate
competitively from a single initial  phase. For such a
system, we present a general method to extract from numerical
simulations the individual nucleation rates of the nucleating phases.
The results show that the Ostwald rule, which predicts which phase
will nucleate, must be modified probabilistically when the new phases
are almost equally stable. Finally, we show that the nucleation rate
of a phase depends, among other things, on the number of other phases
accessible from it.
\end{abstract}

\date{\today}

\maketitle

Metastability is a ubiquitous phenomenon in nature. Broadly speaking,
it occurs when a system is ``trapped'' in a phase different from equilibrium.
This non-equilibrium phase, the metastable state, can last for extremely
long times.  Thus, it is not surprising that metastable
states play a crucial role in many physical processes and are at the
center of much current research. For example, recently an intermediate
metastable phase was shown to provide an easier pathway for the growth
of crystal nuclei from fluids (\emph{nucleation}), with implications
for the crystallization of proteins \cite{LutskoNicolisPRL2006,
tenWoldeFrenkelProteinNucleationScience1997}. Proteins
themselves are known to get stuck in misfolded metastable structures
\cite{WolynesMetastableProteinPRE1997}, preventing them from reaching
their equilibrium configuration.  The phenomenology observed in these
and in many other systems can be thought of as arising from a
complicated energy landscape, with several local ``metastable minima''
where the trapping occurs \cite{WalesBook}.  The extreme situation
is that of glasses, in which the energy landscape can have
extemely many local minima hindering relaxation of the system to a
thermodynamically stable crystal
\cite{BiroliKurchanMetastableGlassy2001}.

The above systems present at least several metastable
states.  These states and the transitions between
them usually arise from the microscopic interactions in a complicated way. 
When this is the case, the
study of phenomena such as competition between nucleating phases and specific
nucleation pathways may be
obscured. In view of this, in this work we present a simple spin model
with nearest-neighbour interactions, where the number of metastable
phases and the decay pathways between them can be explicitly specified
by varying the model parameters. It thus serves as a test-bed for
theoretical results relating to systems with multiple metastable
phases \cite{BovierCMP2002,GaveauSchulmanMultiplePhasesPRE2006,
LarraldeLeyvrazSandersJStatMech2006}, just as the
kinetic Ising model, a special case of our model, has been central
in the study of systems with a single metastable phase
\cite{RikvoldLifetimes1994}. As discussed below, the model also describes the
adsorption of multiple chemical species onto a surface, an
interesting physical problem in its own right.

After presenting the model, as an illustration 
of a possible application, 
we construct a system with arbitrarily long
successions of metastable states. We then focus on competition between
phases nucleating from a single initial metastable phase. An important
question in this context is to understand which phases nucleate under
which conditions. The Ostwald rule states that the nucleating phase is
the one with the smallest free energy barrier from the
initial phase: see Ref.~\cite{FrenkelOstwaldStepRulePCCP1999} and
references therein. Previous results have supported this prediction
\cite{SearImpurities2005}.

We show that in general the Ostwald rule must be modified
probabilistically when the new phases are of similar stability, using
an argument based on individual nucleation rates of each phase.  We
give a method by which these rates can be measured in simulations or
experiments, and show that there is a parameter regime in which any of
the new phases may nucleate---only the nucleation probability of each
phase can be established, with the outcome in any given run being
unpredictable. We finally show that the nucleation probability of a
phase depends on the phases accessible from it.

\paragraph{Model details:-}
Our model is based on the Potts model, in which each spin has one of
$q$ states \cite{WuPottsModel1982} and each phase has a
majority of spins in one state; the Ising model corresponds to
$q=2$.  The relative stability of each phase is controlled by external
fields, and the interplay of these fields with interactions between
different spin states allows us to obtain any desired transition
pathways between phases.

Viewing the fields as chemical potentials, we can recast the model as a
multi-component lattice gas which describes adsorption on a lattice
substrate (e.g., a crystal plane) of multiple chemical species with
lateral interactions
\cite{RikvoldLatticeGasMulticptAdsorptionSurfSci1988}.  Much
experimental work has been done on the thermodynamics of such systems,
but little on the kinetics---see
\cite{PereyraAdsorptionMulticptLangmuir2004} and references therein;
nonetheless, our results should be testable in that context.  A more
complicated system where the kinetics has been characterized is a
colloid--polymer system \cite{PoonEvansColloidPolymerTripleCoexPRL1999,
PoonEvansKineticsColloidPolymerTripleCoexPRE2001}, where
possible pathways were found from considerations of the free energy
landscape
\cite{PoonEvansClassificationOrderingKinetics3PhasePRE2001}.
Our approach is complementary in that specific pathways result from
microscopic interactions. 

We work on an $L \times L$ square lattice with $N\defeq L^2$ spins 
and periodic boundary conditions, although the results are qualititavely 
unaffected by lattice type. 
Each lattice site $i$ has a spin $\s_i$ taking values in $\{1, \ldots,
q\}$, and the energy of a configuration $\ss$ is given by the
Hamiltonian
\begin{equation}\label{eq:hamiltonian}
H(\ss) 	\defeq -\sum_{\langle i, j \rangle} \J{\s_i}{\s_j} -
\sum_{\alpha=1}^q h_{\alpha} M_{\alpha}.
\end{equation}
Here, $M_{\alpha} \defeq \sum_{i} \delta_{\s_i,
\alpha}$
is the magnetisation ($=$number of spins) of the
spin type $\alpha$;
$\delta_{\alpha, \gamma}=1$ if $\alpha = \gamma$, and
$0$ otherwise.  The first term is a sum over nearest-neighbor pairs of
spins of a symmetric interaction energy
$\J{\alpha}{\gamma} = \J{\gamma}{\alpha}$, and the second describes
the effect of external fields $h_{\alpha}$ acting on spin type
$\alpha$.

We set the diagonal elements $\J{\alpha}{\alpha}$ of the interaction
matrix to unity ($\J{\alpha}{\alpha} \defeq 1$ for all $\alpha$), so
that in the absence of non-diagonal interactions and fields, the model
reduces to the standard Potts model \cite{WuPottsModel1982}. This has
$q$ symmetrical phases coexisting below a critical temperature $T_c =
1/(\ln(1+\sqrt{q}))$; each phase has a majority of spins in one of the
$q$ spin states.  Including fields breaks the symmetry between
phases. If $h_\alpha = h_\gamma$, then the $\alpha$ and $\gamma$
phases coexist, with a first-order phase transition between them, for
$T < T_c$; this is at the origin of metastability.  Weak non-diagonal
interactions do not qualititavely affect this coexistence.

To evolve the system we choose discrete-time Metropolis dynamics
\cite{NewmanBarkemaBook}: at each time step, a spin and its new value
are chosen at random, the increment
$\Delta H$ of the Hamiltonian \eqref{eq:hamiltonian} for this change
is calculated, and the update is accepted with probability $\min\{1,
\exp(-\beta \Delta H)\}$, where $\beta := 1/T$ is the inverse
temperature. This gives a Markov chain on the space of all possible
configurations.

This Markov chain has a unique equilibrium distribution, concentrated
on the phase(s) with the largest $h_\alpha$.  The other phases are
\emph{metastable}, that is, when started in such a phase $\alpha$, the
system stays there for some time, before a transition to a more stable
phase $\gamma$ is \emph{nucleated} by the appearance of a critical
droplet of the $\gamma$ phase.  At sufficiently low temperatures, the
relative stability is determined by $h_\gamma > h_\alpha$.  The
reverse transition is exponentially unlikely.

In the standard Potts model, the equilibrium phase (almost) always
 nucleates.  To obtain non-trivial transition pathways, nucleation of
other phases must be promoted.  This we achieve using
 non-diagonal interactions between distinct spin types $\alpha \neq
\gamma$: setting $\J{\alpha}{\gamma}>0$ favors nucleation of $\gamma$
droplets inside the $\alpha$ phase by lowering the surface tension
between $\alpha$ and $\gamma$ regions, and hence decreasing the
droplet free energy of formation (\emph{nucleation barrier}),
whereas formation of $\gamma$ droplets in the $\alpha$ phase is
suppressed if $\J{\alpha}{\gamma}<0$.

\begin{figure}
\includegraphics{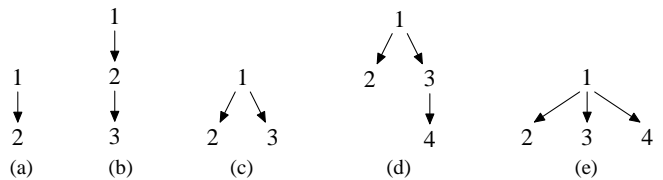}
\caption{Metastable transition graphs: (a)~kinetic Ising model;
(b)~succession of $3$ phases; (c)~single metastable phase decaying to
two competing phases; (d)~as in (c), but such that one phase can decay
further; (e)~three competing phases.}
\label{fig:transitions}
\end{figure}

We can now construct models whose phases obey arbitrary
\emph{metastable transition graphs}. These are directed graphs with
the restriction that no loops returning to a previously visited phase
are allowed.  Each vertex corresponds to one phase, labeled by its
dominant spin state, and each arrow to a desired transition: $\alpha
\to \gamma$ means that phase $\gamma$ can nucleate directly from phase
$\alpha$. \figref{fig:transitions} shows example transition
graphs. 

To construct a model corresponding to a given transition graph, we proceed as
follows. The number of spin types, $q$, is the number of
vertices in the graph. To each spin type $\alpha$ we assign a field
$h_{\alpha}$, with $h_{\gamma} > h_{\alpha}$ if $\gamma$ is below
$\alpha$ in the graph.  The off-diagonal interactions are given by
$\J{\alpha}{\gamma} \defeq K_1 > 0$ (attractive) if $\alpha \rightarrow \gamma$,
and $\J{\alpha}{\gamma} \defeq -K_2 < 0$ (repulsive) otherwise.
$K_2$ must be large enough to inhibit
immediate formation of non-adjacent phases with large fields.

\begin{figure}
\includegraphics{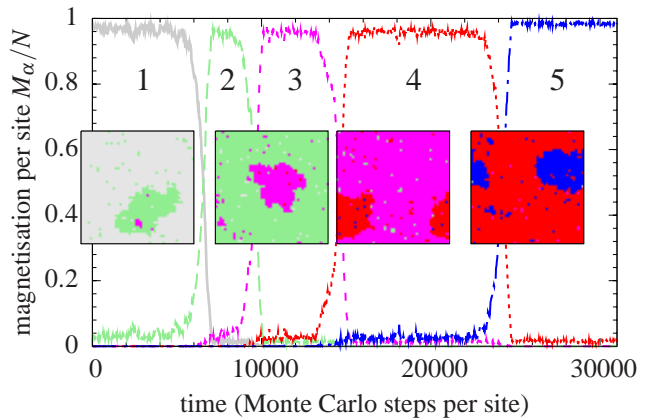}
\caption{(Color online) Time dependence of magnetisation per site
$M_\alpha / N$ of each phase $\alpha$, in a single run of the model
with transition graph $1 \rightarrow 2 \rightarrow 3 \rightarrow 4
\rightarrow 5$ (a succession of phases). Parameters are $L=50$, $\beta
= 1.25$, $h_{\alpha} = 0.1(\alpha -1)$, $K_1 = 0.1$, and $K_2 =
1.0$. Labels denote the dominant phase. Configuration snapshots depict
post-critical nuclei of each new phase embedded in the previous
phase. These grow to fill the system, producing the next phase in
sequence.
}
\label{fig:succession}
\end{figure}

As an illustration, we construct a model exhibiting a linear
succession of metastable phases with transition graph $1 \to 2 \to
\cdots \to q$. We impose fields $0 = h_1 < h_2 < \cdots < h_q$ and
attractive interactions $\J{\alpha}{\alpha\pm1} \defeq K_1 > 0$
between neighbouring states, and set all other non-diagonal
interactions to $-K_2 < 0$. With suitable, moderately robust,
parameters, we observe the desired behavior, shown for $q=5$ in
\figref{fig:succession}.

A three-phase succession was previously observed in a kinetic
Blume--Capel model \cite{CirilloOlivieriMetastableBlumeCapel1996,
RikvoldNovotnyCompetingMetastablePRE1994}, corresponding to a
special case of our model with $q=3$ \cite{OurUnpublished}. The
physical reason for the observed transitions is, however, much
more transparent with the Hamiltonian in the form
\eqref{eq:hamiltonian}, with its intuitive interpretation in terms of
attractive and repulsive interactions.

\paragraph{Competitive nucleation:-}
We now turn to the decay of one metastable phase into two competing
phases (\figref{fig:transitions}(c)).  Sear studied competitive
heterogeneous nucleation (occurring on impurities) in the $3$-state
standard Potts model \cite{SearImpurities2005}. In contrast, all
behaviors discussed in this work are \emph{endogenous}: observed
transitions are not caused by external influences, but rather arise
spontaneously from within the system itself.

We fix $0=h_1 < h_2 \le h_3$, $\J{1}{2} = \J{1}{3} > 0$ and $\J{2}{3}
< 0$.  Let $\Delta h \defeq h_3 - h_2$ be the field difference
between the new phases, $2$ and $3$.  When $\Delta h=0$, these phases are
symmetrical, each nucleating half of the time, while for $\Delta
h >0$, we expect the $1$--$3$ free energy barrier to be lower than the $1$--$2$
one, so that according to the Ostwald rule, only phase $3$ should
nucleate.
To test this, we perform $n$ simulations starting from phase $1$ for
each $\Delta h$, in $n_2$ of which phase $2$ nucleates before phase
$3$.  The ratio $n_2 / n$ is then an estimate of the probability
$p_2(\Delta h)$ that phase $2$ nucleates first.  For efficiency, we
use a rejection-free version of the Metropolis method
\cite{NovotnyRejectionFree1995,OurUnpublished}.

\begin{figure}
\includegraphics{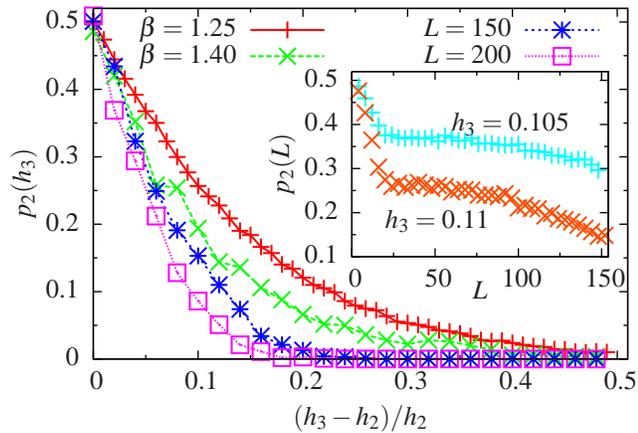}
\caption{(Color online) Probability $p_2(h_3)$ that phase $2$
nucleates before phase $3$ as a function of $\Delta h/h_2$, with
$h_2=0.1$, $K_1 = 0.1$ and $K_2=1$, and $\beta=1.25$, $L=50$ unless
otherwise noted.  Up to $n=10^4$ trials were used for each data point;
statistical errors are of the order of the symbol size.  Inset:
system-size dependence of $p_2$ for two values of $h_3$ with fixed
parameter values.}
\label{fig:nucleation-prob}
\end{figure}

\figref{fig:nucleation-prob} plots $p_2$ as a function of a
non-dimensionalised $\Delta h$. For $\Delta h$ sufficiently close to
$0$, phase $2$ can still nucleate first, contrary to the simple
Ostwald rule. The probability that it does so rapidly decreases for
larger $\Delta h$, until a point beyond which phase $3$ effectively
always nucleates.

To explain these results in a general context, we assume, as in
classical nucleation theory \cite{DebenedettiBook},
that there are well-defined nucleation rates $\lambda_i(L)$ of phases
$i=2,3$, giving the number of critical nuclei which form
per unit time in a system of size $L$. The nucleation rates per site
are $\mu_i(L) \defeq \lambda_i(L) / N$. 

A nucleation rate is the inverse of a mean nucleation time, which can
be measured in experiments or simulations by averaging over many
nucleation events in independent runs. In the case of competitive
nucleation, however, we can only measure the mean time $\tau$ for the
\emph{first} phase to nucleate, after which this phase invades the
entire system.  The rate of this first nucleation is $\lambda_2 +
\lambda_3$, since the total number of nucleation events per unit time
is the sum of those for each type, so that $\tau = 1/(\lambda_2 +
\lambda_3)$.  For convenience, in simulations $\tau$ is taken to be
the time until the new phase occupies half the system.

Under the same assumptions, the probability that phase $2$ nucleates
first is $p_2 = \Prob{T_2 \le T_3}$, where $T_i$, the time for phase
$i$ to nucleate, is an exponentially distributed random variable with
mean $1/\lambda_i$.  This gives $p_2 = \lambda_2/(\lambda_2 +
\lambda_3) = \lambda_2 \tau$.

Individual nucleation rates of the two phases, which \emph{cannot} be
obtained directly, can now be calculated as: $\lambda_2 = p_2 / \tau$ and
$\lambda_3 = (1-p_2) / \tau$. This generalizes to $P$
competing phases, where the measurable quantities are the mean
nucleation time $\tau = 1/\sum_{i=1}^P \lambda_i$ and the nucleation
probabilities $p_i = \lambda_i / \sum_j \lambda_j = \mu_i / \sum_j
\mu_j$ of each phase $i$. The nucleation rate of phase $i$ is then
$\lambda_i = p_i / \tau$.

\begin{figure}
\includegraphics{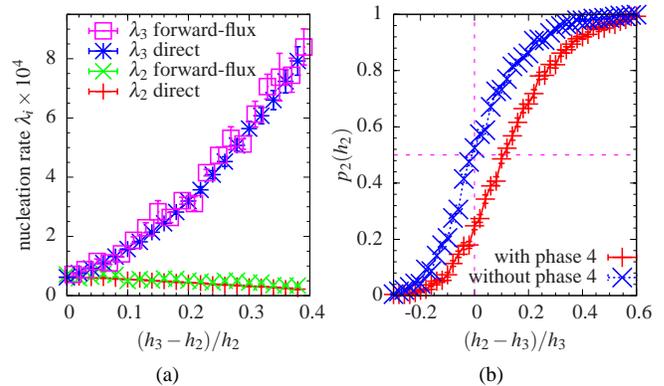}
\caption{(Color online) (a) Comparison of nucleation rates
$\lambda_i(h_3)$ of phases $i=2,3$ from phase $1$ in the model of
\figref{fig:transitions}(c) with $h_2$ fixed, calculated directly from
simulations and using forward-flux sampling. (b) Nucleation
probability $p_2(h_2)$ for $h_3=0.1$ and $h_4 = 0.2$ varying $h_2$,
with and without phase $4$, in the model of
\figref{fig:transitions}(d). Dashed lines show positions of equal
nucleation probability and equal field of the two phases. In both
subfigures parameters are as in \figref{fig:nucleation-prob}, with
$L=50$ and $\beta=1.25$.}
\label{fig:nucleation-rates}
\end{figure}

\bfigref{fig:nucleation-rates}(a) shows $\lambda_2$ and $\lambda_3$ calculated
in this way.  To confirm the
validity of such calculations, we use the
``forward-flux sampling'' method
\cite{AllenSwitchingBiochemicalForwardFluxPRL2005,
AllenRareEventsStochasticJCP2006}, which directly calculates
 the transition rate between two phases in a stochastic system.
This has previously been used to study nucleation rates in the Ising model
\cite{SearHetHomNucleationComparedJPCB2006,
SearHeterogeneousNucleationPoresPRL2006}.  In our case, the possibility of
escape to an additional new phase must be taken into
account \cite{OurUnpublished}.
\bfigref{fig:nucleation-rates}(a) shows that the results indeed coincide 
with those of the direct method, within
statistical errors.
We remark that we are unaware of any
analytical prediction giving the observed variation of nucleation rates.

The above considerations used ``in reverse'' confirm that the Ostwald
rule must in general be modified when the new phases have similar stabilities,
as follows. Consider phases which are equally stable for given parameter
values.  We expect nucleation barriers, and hence nucleation rates, to
vary continuously with the parameters, so that the nucleation
probabilities $p_i$ also vary continuously. Hence there is a region,
where the phases have similar stabilities, in which all nucleation
probabilities are non-zero---only the probability of each phase
nucleating is well-defined, with the outcome in any given run being
stochastic, as in \figref{fig:nucleation-prob}.  The definite prediction given
by the Ostwald rule is thus invalid in this region.

To see how our results depend on system size $L$, we note that in a broad
region of $L$, the per-site nucleation rates
$\mu_i$, and hence also the $p_i$, are \emph{independent} of $L$, as
confirmed by the plateaus in the inset of
\figref{fig:nucleation-prob}. This is valid when 
the nucleation process is mediated by growth of a single droplet
\cite{RikvoldLifetimes1994}. Note that this regime may be of relevance
for macroscopically large systems \cite{RikvoldLifetimes1994}.

\begin{figure}
\includegraphics{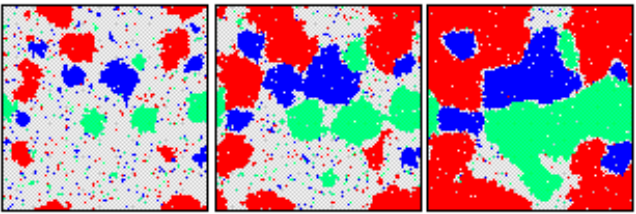}
\caption{(Color online) Configuration snapshots of the
multidroplet regime in the model with transition graph
\figref{fig:transitions}(e), with
$L=200$, $\beta=1.25$, and $h_1=0$, $h_2=h_3=h_4=0.15$. 
Nucleation of droplets of three equally stable phases from a single metastable
phase is followed by droplet growth and then domain coarsening.  The
``coating'' of the initial phase visible between domains in the final snapshot
is due to repulsive interactions between the new phases.}
\label{fig:multi-droplet} 
\end{figure}

Above a certain system size, however,
 droplets of different phases may nucleate before any
dominates the system (the `multidroplet' regime) \cite{RikvoldLifetimes1994}.
This results in
coarsening, as shown in \figref{fig:multi-droplet} for three competing
phases of equal stability.  Even if a phase-$\alpha$ droplet nucleates first,
droplets of a more-stable phase may then nucleate and grow to dominate the
system before the $\alpha$ phase can do so, thus reducing $p_\alpha$,
as seen in
\figref{fig:nucleation-prob}. 

Finally, a generic non-symmetric case can be obtained by adding a new
phase, $4$, and a decay path, $3 \to 4$, as in
\figref{fig:transitions}(d).  Even when $h_2=h_3$, phase $3$ now nucleates
more often, as shown by the horizontally displaced 
nucleation probability curve in \figref{fig:nucleation-rates}(b):
the presence of phase $4$ reduces $p_2$ by roughly half. This is due
to an entropic effect: there are more $3$-dominated critical droplets
than $2$-dominated ones, since spins of type $4$ also appear in the
former droplets, resulting in nucleation of a binary mixture
\cite{DebenedettiBook}; a similar effect is visible in
\figref{fig:transitions}. There is thus a lower free energy barrier to
form phase $3$, and yet nucleation of both phases $2$ and $3$ is still
observed, again at odds with the simple Ostwald rule.

In summary, we have introduced a generalized Potts model which can
easily be tuned to have any given number of metastable phases and
arbitrary transitions between them. We have shown generally that
individual nucleation rates of competitively nucleating phases can be
calculated from experimentally measurable quantities, and that the Ostwald
rule must be modified when the nucleating phases have comparable stabilities.
 In future work
\cite{OurUnpublished}, we will study the model in detail and compare
its properties with theoretical results
\cite{LarraldeLeyvrazSandersJStatMech2006}
on systems with multiple metastable phases.  

DPS thanks A.~Huerta for useful discussions and the Universidad
Nacional Aut\'onoma de M\'exico for financial support.  The financial
support of DGAPA-UNAM project PAPIIT IN112307 is also acknowledged.


\end{document}